\documentclass[
prl
,showpacs
,showkeys
,aps
,amsfonts
,amssymb
,byrevtex
,pdftex  
,twocolumn
]{revtex4-1}

\usepackage{graphicx}  
\usepackage{amsmath, amssymb, bm}   
\usepackage[dvips]{color}
\usepackage{natbib}
\usepackage{hyperref}

\def\r{{\it r}}

\def\){\right)} 
\def\({\left(} 
\def\]{\right]} 
\def\[{\left[}

\begin{document}

\title{
The  \r-mode instability  in strange stars with a crystalline crust}

\author{%
Gautam Rupak}
\email{grupak@u.washington.edu}

\affiliation{Department of Physics $\&$ Astronomy, 
Mississippi State
University, Mississippi State, MS 39762, U.S.A. }

\author{%
Prashanth Jaikumar
}
\email{pjaikuma@csulb.edu}

\affiliation{Department of Physics $\&$ Astronomy,
California State University Long Beach, Long Beach, CA
  90840, U.S.A. }
\begin{abstract}

The \r-mode instability, believed to limit the rotation speed of compact stars, can provide empirical confirmation for the existence of stable deconfined phases of quark matter that are predicted by weak coupling calculations in Quantum Chromodynamics. We construct a model for strange quark stars as heavy as 2$M_{\odot}$ that are made of superconducting quark matter in the bulk and a thin crystalline quark matter crust. This crystalline quark crust is sufficiently robust to withstand \r-mode heating and viscous rubbing for realistic mode amplitudes ${\cal O}(10^{-2})$, unlike a crust made of neutron-rich nuclei. The dissipation provided by viscous rubbing at the core-crust boundary is both necessary and sufficient to obtain stable rotation speeds that are consistent with the majority of rapidly spinning pulsars in low mass X-ray binaries. Our analysis implies that while bare strange stars are ruled out by the existence of rapidly spinning pulsars,  a strange star with a quark matter crust is a distinct possibility.

\end{abstract}



\maketitle

\emph{Introduction.}--- 
Compact stars provide a unique window into the physics of matter at densities well above nuclear saturation.  At asymptotically high densities,  Quantum Chromodynamics (QCD) favors a maximally symmetric phase of superconducting quark matter called the color-flavor-locked (CFL) phase~\cite{Alford98}. However, at compact star densities, the phase of matter is less certain. In this intermediate density regime, some of the most important constraints on theoretical models of dense matter come from compact star observables,  eg., mass-radius relations~\cite{Steiner:2010fz} and cooling rates in compact stars~\cite{Page:2010aw}. 
The spin frequencies of fast spinning pulsars is yet another, but herein lies a puzzle.
 
 The spin frequency $\nu\approx 716$ Hz of the fastest spinning neutron star PSR J1748-2446ad~\cite{Hessels} is still well below the estimated theoretical maximum $\approx 1000$ Hz corresponding to the break-up Kepler frequency. The majority of ``spun-up'' neutron stars in low mass X-ray binaries (LMXBs) cluster between 300-640 Hz~\cite{Chakrabarty:2008gz}. What determines the maximum spin frequency and how does it relate to the nature of dense matter inside the star? 
 It is believed that the $r$-mode instability~\cite{Andy,morsink} sets the limit on spin frequency. 
 The \r-mode is driven unstable by gravitational wave emission at the expense of rotational energy, thereby spinning the star down. Internal sources of friction such as bulk/shear viscosity  tend to damp $r$-mode growth.   At any given temperature $T$, the competition between the former growing on a characteristic time scale $\tau_\mathrm{GW}$ and the latter acting on a time scale  $\tau_\mathrm{F}$ determines the limiting or critical frequency $\Omega_c$.   
  The gravitational waves from $r$-mode oscillations 
may be detected given the expected improved sensitivities of 
ground-based interferometers (eg. VIRGO, Advanced
LIGO~\cite{Sa,Watts}), providing a novel diagnostic of dense matter in compact stars. 

In this letter, we investigate the maximum spin frequency of strange stars that can sustain the \r-mode instability. If discovered, strange stars, which are believed to be made entirely of deconfined quarks, would confirm Witten's hypothesis of self-bound strange quark matter~\cite{1984PhRvD..30..272W} as the true ground state of matter. At the theoretical level, the \r-mode mechanism for limiting the star's spin frequency does not seem to work for ${\it bare}$ strange stars, i.e, one without a crust~\cite{Madsen00,Kokko,JRS08,Man,Rupak:2010qg}. The critical frequency falls short of typical LMXB frequencies, essentially because of insufficient dissipation of the \r-mode energy. However, studies of ordinary neutron stars~\cite{Levin:2000vq, Lindblom:2000gu, Lee:1996rx} suggest that \r-modes can be strongly affected (damped) by the presence of a nuclear crust.
This extra dissipation compared to a pure fluid is mediated by kinetic viscosity in a thin boundary layer and significantly shortens $\tau_\mathrm{F}$. Consequently, the critical frequency $\Omega_c$ increases dramatically. Such a large effect is not expected for neutron stars because even \r-modes with relatively small amplitude ${\cal O}(10^{-2})$ can melt the nuclear crust~\cite{Lindblom:2000gu}. But compared to a nuclear crust, the crystalline quark phase is much more rigid, with a shear modulus that is $20-1000$ larger~\cite{Rajagopal:2006dp} and a melting temperature $T_{\rm melt}\approx \Delta\approx 10$ MeV that is 10 times larger. If crystalline quark matter, rather than nuclei, is taken to constitute the crust, it allows for a large damping of the $r$-mode that brings the strange star model into agreement with observed LMXB spin rates. 
We find that if one is to entertain the Witten hypothesis of the absolute stability of quark matter at high density, self-bound strange stars  likely have a crust made of superconducting quarks in the crystalline phase.

\emph{Self-bound stars with a crystalline crust.}---
At realistic compact star densities with quark chemical potential $\mu_q\sim 300$ MeV, the strange quark mass $m_s\sim 150$ MeV is non-negligible, and severely stresses the CFL phase. The energetically favored phase is a chirally rotated phase of CFL~\cite{Bedaque:2001je} which leads to kaon condensation when the energy cost $m_s^2/(2\mu_q)$ exceeds the CFL kaon mass $m_K$. A homogeneous strange star in this kaon condensed CFL phase (henceforth CFL-K0) has already been ruled out from an $r$-mode study~\cite{Rupak:2010qg}. We consider the effect of a thin layer of superconducting quark matter in the crystalline phase 
forming a crust on the surface of a CFL-K0 star. In the CFL-K0 phase, the equation of state  can be written as~\cite{Alford:2004pf}  
\begin{align}
P_\mathrm{CFL-K0}=&\frac{1}{3}(\epsilon-4 B)+\frac{4\Delta^2-m_s^2}{3\pi}\sqrt{\frac{\epsilon-B}{a_4}}\nonumber\\
&-\frac{m_s^4}{12\pi^2}\[1-2\pi^2\bar{f}_\pi^2 + \frac{(4\Delta^2-m_s^2)^2}{a_4 m_s^4} \right. \nonumber\\
&\left.-3\ln\frac{8\pi}{3m_s^2}\sqrt{\frac{\epsilon-B}{a_4}}\] , 
\end{align}
where $B$ is the MIT Bag constant, the decay constant $\bar{f}_\pi^2=(21-8\ln2)/(36\pi^2)$, and $a_4\sim 0.7$ is a reasonable choice for perturbative QCD-inspired corrections~\cite{Fraga:2001id}. 
This parameterization allows for quark stars as heavy as 2$M_{\odot}$ and a phase transition to quark matter above nuclear saturation. For a fixed value of the strange quark mass $m_s$, CFL-K0 gap parameter $\Delta$ and $a_4$,  the bag constant $B$ is determined by our choice of a given binding energy per baryon $E/A$ for bulk strange quark matter. 
The mass $M$  and radius  $R$ of the compact star is calculated using the Tolman-Oppenheimer-Volkov equations with general relativistic corrections. For fixed $m_s$, $\Delta$, $a_4$  and $B$, the central density $\rho_c$ of the star is adjusted as a factor of the nuclear saturation density  $\rho_0=2.55\times 10^{14}\ \mathrm{g/cm^3}$ to produce different stellar masses.  The core-crust boundary is determined by comparing the free energy of the CFL-K0 with the most favored crystalline quark phase at these densities~\footnote{
The gapless CFL phase could possibly intervene between the CFL-K0 and the crystalline phase~\cite{Alford:2003fq}. However, it has a chromomagnetic instability~\cite{Huang:2004bg}, with complicating features, so we have chosen a simpler case.}. 
This is the CubeX phase, a specific generalization of the LOFF phase where  the  quark pairs carry non-zero total momentum 2${\bm q}_n$~\cite{Rajagopal:2006dp}. 
The gap varies in space as $\sim {\rm exp}(i{\bm q}_n\cdot{\bm x})$ according to the reciprocal vectors $\{\bm{q}_n\}$ which define the specific crystal structure of this phase (as opposed to a physically rigid lattice)~\footnote{For details of the gap calculation and free energy, the reader is referred to Eq.(74) and Sec. E of~\cite{Rajagopal:2006dp}.}. The chemical potential at which the crystalline phase takes over is quite sensitive to the gap, hence we tune the gap to locate this transition point near the surface of the star at radius $r_c< R$. Though the gap, and hence the location of the crust cannot be known for certain, the above construction provides the most instructive scenario, in that it allows us to demonstrate the effect that even a thin crust $d_{\rm crust}= R-r_c\ll R$ can have in providing additional damping compared to homogeneous quark phases.  Table~\ref{Table:TOV} shows the two sets of parameters we use in our calculations. 
\onecolumngrid
\begin{widetext}
\begin{table}[h]
\centering
\begin{ruledtabular}
\begin{tabular}{c c c c c c c c c c c}
$m_s$ (MeV) & $B^{1/4}$ (MeV) & $E/A$ (MeV) & $a_4$ & $\Delta$ (MeV)&$\rho_c (\rho_0)$&$R$ (km)&$M (M_\odot) $&$r_c (R)$&$\mu(r_c)$ (MeV)& $\mu(R)$ (MeV) \\ \hline
160& 137.8 &  930& 0.7& 24.0& 2.71 & 11.14 & 1.40 & 0.997 & 310.3 & 310.0\\
160& 136.1 &  920& 0.7& 24.2& 4.93 & 11.71 & 1.97 & 0.993 & 307.7 & 306.7
\end{tabular}
\end{ruledtabular}
\caption{\protect Parameters for two self-bound stars, $M/M_{\odot}$=1.40 and 1.97. The location of the crust in stellar units is $r_c$, whereat the transition chemical potential is $\mu(r_c)$.}
\label{Table:TOV}
\end{table}
\end{widetext}
\twocolumngrid


\emph{$r$-mode damping and critical spin frequency.}--- 
The restoring force for the \r-mode oscillation is the Coriolis force, and ignoring viscous effects, the mode frequency $\omega_r$ in the inertial frame is~\cite{morsink}
\begin{align}
\label{kappa2} 
\omega_r &=  \omega_{\rm rot}-m\Omega
\approx \[\frac{2}{m+1}-m\]\Omega 
+\mathcal O(\Omega^2)\, ,
\end{align}
where $\omega_{\rm rot}$ is the mode frequency in the frame that co-rotates with the star, and $m$=$2$ is the azimuthal quantum number of the first \r-mode to become unstable. Higher order corrections include the sensitivity to the density profile $\rho(r)$ (see Refs.~\cite{LMO,JRS08}). 
The \r-mode energy
grows at the expense of the star's rotational kinetic energy as angular momentum is lost to gravitational wave emission. In general, viscous forces counter this energy growth. 
The time scale $\tau$ associated with growth or dissipation is given by $1/\tau_i= -\frac{1}{2E}\(\frac{dE}{d t}\)_i$, where the subscript corresponds to the growth or dissipative mechanisms and $E$ is the mode energy.  The time scale for gravitational radiation is given by~\cite{morsink}
\begin{align}
\label{taugw}
\frac{1}{\tau_\mathrm{GW}} = - \frac{32 \pi G}{c} 
\frac{\left(m-1\right)^{2m}}{\left[\left(2m+1\right)!!\right]^2} 
\int_0^R dr\rho(r)\[r \frac{\Omega}{c}
  \frac{m+2}{m+1}\]^{2m+2}\,  .
\end{align}
This timescale is negative, indicating exponential
mode growth (the instability). The damping timescales from shear viscosity $\tau_\eta$ and bulk viscosity $\tau_\zeta$ for the CFL-K0 phase have been calculated in Ref.~\cite{Rupak:2010qg}. The important new contribution comes from the damping in the viscous boundary layer (VBL) just beneath the rigid crust located at $r_c$ (Table \ref{Table:TOV}). The $r$-mode oscillations decay in the crust exponentially within a characteristic length scale~\cite{Lindblom:2000gu}
\begin{align}
\label{rootvisc}
d=\sqrt{\frac{\eta(r_c)}{2\Omega\rho(r_c)}}\,  ,
\end{align}
where $\eta(r_c)$ is the shear viscosity of the matter at density $\rho(r_c)$. For the range of parameters we explored, the crust thickness $R-r_c$ is always at least a factor of about 100 larger than $d$, so the VBL approximation is self-consistent. The damping time associated with dissipation in the VBL is  called the rubbing time scale, given by~\cite{Lindblom:2000gu}
\begin{align}
\tau_\mathrm{rub} =\frac{2^{m+2}(m+1)!}{m(2m+1)!! \operatorname{I}_m}\sqrt{\frac{\rho(r_c)}{\Omega\eta(r_c)}}
\int_0^{r_c} dr \frac{r^{2m+2}\rho(r) }{r_c^{2m+2}\rho(r_c)}\,  ,
\end{align}
where $I_m\approx 0.8041$. 
The above estimate assumes that mechanical coupling between the core and the crust is  absent, but if the $r$-mode oscillations penetrate into the crust, it increases the rubbing time scale by a  ``slippage" factor of $(\delta u/u)^{-2}$~\cite{Levin:2000vq,Lindblom:2000gu}, where $u$ is the oscillating fluid velocity in the core and $\delta u$ the difference in the fluid velocity in the core and the crust across the boundary. For a nuclear crust, the slippage factor spans a range $0.003\lesssim (\delta u/u)^{2}\lesssim 1$ over a wide range of spin frequencies $0\leq\Omega/\Omega_K\leq 0.5$.  $\Omega_K
 =\frac{4}{9}\sqrt{2\pi G \bar\rho}$  
is the Kepler frequency with $\bar\rho$ being the average matter density.
 Crystalline quark matter is much more rigid than the nuclear crust. Performing a mode penetration calculation for the quark crust along the lines of Ref.~\cite{Levin:2000vq}, we find that the slippage factor for a canonical $r$-mode frequency $\omega=2\Omega/3$ is nearly 1 for the entire frequency range of interest, assuming a shear modulus about 100 times larger than the nuclear crust. This is reasonable given how rigid the quark crust is.

\begin{figure}[thb]
\begin{center}
\includegraphics[width=0.49\textwidth,clip=true]{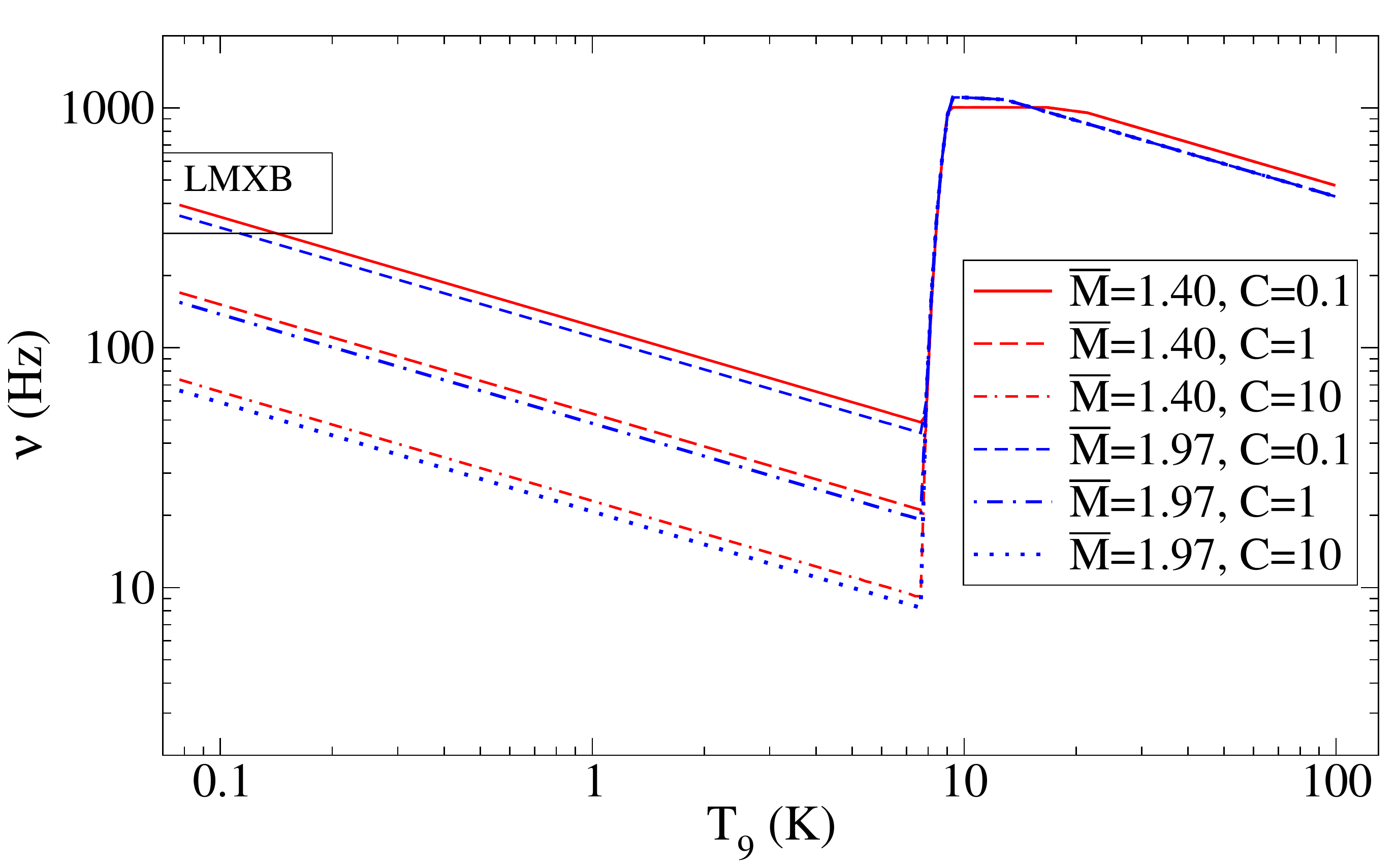} 
\end{center}
\caption{\protect Critical frequency for the $\overline{M}=M/M_\odot =1.97$ star (blue curve) and $\overline{M}=M/M_\odot=1.4$ star (red curve).}
\label{fig:CriticalCurveB}
\end{figure}

The critical frequency of the star at any temperature can be determined by the criterion that at this frequency, the fraction of energy dissipated per unit time by viscous processes exactly cancels the fraction of energy fed per unit time into the $r$-mode by
gravitational wave emission:
\begin{align}
\label{criteqn} \frac{1}{\tau_{\rm total}}\Big|_{\Omega_c}\equiv\left[\frac{1}{\tau_{\rm GW}}+ \frac{1}{\tau_{\eta}}+\frac{1}{\tau_{\zeta}}+\frac{1}{\tau_{\rm rub}}\right]\Big|_{\Omega_c}=0\,  . 
\end{align}
The critical frequency curves obtained from solving Eq.~(\ref{criteqn}) are plotted in Fig.~\ref{fig:CriticalCurveB}, where at frequencies below  the curve the  $r$-mode oscillations are suppressed  due to the dissipative processes.  We choose parameters for a small kaon condensation such that $\delta m\equiv m_K-(m_s^2-m_d^2)/(2\mu_q)=-2$ MeV, with down quark mass $m_d=7$ MeV. 
The kaon interaction coupling $C$ is varied over a reasonable range $0.1$ to $10$~\cite{Alford:2009jm}.  Weaker coupling gives larger shear viscosity. 
The shear viscosity is sensitive to the kaon condensation, increasing sharply with decreasing $\delta m$~\cite{Alford:2009jm}. 
However, one cannot decrease the coupling $C$ or $\delta m$  arbitrarily 
to increase the critical frequency and explain the LMXB spin rates. 
Too weak a coupling or small $\delta m$ would make the mean free path of the kaons larger than the star radius invalidating the hydrodynamic calculation.  For example, the dip in the critical frequency around $7\times 10^9$ K corresponds to excluding phonon contributions when its mean free path becomes comparable to the star radius.   
We have shown that in the absence of the crust, even a small $C$ is not sufficient to reproduce LMXB spin frequencies~\cite{Rupak:2010qg}.   We also limit the critical frequency to be below the limit set by the Kepler frequency $\Omega_K$.  This results in the plateaus near $T=10^{10}$ K.  We find in Fig.~\ref{fig:CriticalCurveB} that  at $\delta m\sim -2$ MeV, a small  $C\sim 0.1$ is  compatible with the data. 

Including only the dissipation time from the VBL $\tau_\mathrm{rub}$  and equating it to the gravitational time scale $\tau_\mathrm{GW}$ gives the critical frequency
\begin{align}
\nu_c\approx 123\times T_9^{-5/11}\ \mathrm{Hz}\, , 
\end{align}
where we used parameters for the $1.4 M_\odot$ star with $C=0.1$.  
At $T\approx 10^8$ K, $\nu_c\approx 350$ Hz, thus we see that rubbing friction alone can generate sufficient damping to explain typical
LMXB frequencies. In contrast, equating $\tau_\mathrm{GW}$ to $\tau_\eta$ at $T=10^8$ K gives $\nu_c\approx 11$ Hz. 
These estimates clearly highlight the essential role of the quark crust in providing a large damping of the \r-mode.


\emph{Discussion.}--- 
\r-mode damping in a self-bound strange quark star with a crystalline quark crust is examined. For a plausible range of crust shear modulus, the canonical \r-mode is damped at the core-crust boundary in exactly the right amount to tame the instability at observed LMXB frequencies and temperatures. Additional damping agents, such as mutual friction associated with the phonons and the kaon-condensed phase could provide some extra  small dissipation, but the critical spin frequency estimates presented here are a robust lower bound. Further, we use an equation of state that meets the observational constraint of a large maximum mass of 2$M_{\odot}$, making it a reasonable starting point for further investigation into properties of self-bound stars. Such stars
most likely should have a thin crust made of superconducting quarks in the crystalline phase if the Witten hypothesis of the absolute stability of quark matter at high density holds true. Finally, physical applications of our model might come from analyzing shear modes of the crystalline quark crust and matching them to  frequencies extracted from seismic vibrations in magnetar flares, which appear to pose a challenge for normal quark matter crusts~\cite{Watts:2006hk}.

\emph{Acknowledgments.}---The authors thank Rishi Sharma for useful discussions. The authors are grateful for partial  support  from the Research Corporation for Scientific Advancement through a Cottrell College Science Award for P.J. and  the U.S. NSF Grant No. PHY-0969378 for G.R.


%

\end{document}